\documentclass[10pt]{iopart}
\usepackage{graphicx}
\usepackage{hyperref}

\begin{document}

\title[A novel planar back-gate design for coupled quantum wells]{A novel planar back-gate design to control the carrier concentrations in GaAs-based double quantum wells}

\author{J Scharnetzky$^1$, J M Meyer$^2$, M Berl$^1$, C Reichl$^1$, L Tiemann$^2$, W Dietsche$^1$\footnote{also at MPI for Solid State Research, Stuttgart, Germany} and W Wegscheider$^1$}

\address{$^1$ Solid State Physics Laboratory, ETH Z\"urich, 8093 Z\"urich, Switzerland}
\address{$^2$ Center for Hybrid Nanostructures, University of Hamburg, 22761 Hamburg, Germany}
\ead{janscha@ethz.ch}
\vspace{10pt}
\begin{indented}
\item[]Januar 2020
\end{indented}

\begin{abstract}
The precise control of a bilayer system consisting of two adjacent two-dimensional electron gases (2DEG) is demonstrated by using a novel planar back-gate approach based on ion implantation. This technique overcomes some common problems of the traditional design like the poor 2DEG mobility and leakage currents between the gate and the quantum well. Both bilayers with and without separate contacts have been prepared and tested. Tuning the electron density in one layer while keeping the second 2DEG at fixed density, one observes a dramatic increase of the carrier concentration. This tunneling resonance, which occurs at equal densities of both layers, demonstrates the separated contacts to each individual layer. In another sample with a smaller tunneling barrier and parallel contacted 2DEGs, the transition from a single 2DEG to a bilayer system is investigated at 50 mK in magnetic fields up to 12 T, showing the gate stability in high magnetic fields and very low temperatures. Transitions into an insulating (Wigner crystal) phase are observed in the individual layers in high fields at filling factors below 1/3. The absence of a fractional quantum Hall liquid at filling factor 1/5 in our structure seems to be a consequence of confining the electrons in quantum wells rather than at interfaces. The observed metal-insulator transitions appear to be nearly unaffected by the presence of the second layer separated by a barrier which is only 3 nm thick. We believe that this planar back-gate design holds great promise to produce controllable bilayers suitable to investigate the exotic (non-abelian) properties of correlated states.
\end{abstract}

%
\vspace{2pc}
\maketitle
\ioptwocol

\section{Introduction}

The advent of bilayers of two-dimensional electron or hole gases made the physics of low-dimensional charge gases much richer as they offer a larger variability for the study of the interaction processes. In addition to the intralayer Coulomb interaction, both tunneling and interlayer Coulomb interaction become relevant. Unusual quantum Hall phenomena like the fractional quantum Hall state at filling factor 1/2 have been found\cite{Suen1992,Eisenstein_1992}. The condensation in a BCS-like state has been observed at total filling 1 of the lowest Landau level\cite{Eisenstein2003,Wie2004} which originates from correlations between the layers. Often, these states were found to be accompanied by insulating, most likely Wigner-crystal (WC) phases.\cite{Manoharan_1996,Shayegan_1996,Wiersma_2006,Ding_2014} 

Additionally, many more high-magnetic field quantum phases than the ones just mentioned have been theoretically proposed but still wait to be discovered experimentally. These include quantum fluids showing a non-abelian behavior\cite{Crepel2019}\cite{Zhu_2016}\cite{Peterson_2015,Geraedts_2015}\cite{Faugno_2019}  and exotic crystalline phases.\cite{Faugno2018}

In this manuscript we report the preparation and characterization of a novel bilayer design which we believe will foster the experimental access to this exciting realm of correlation physics. The crucial modification is the use of ion implantation rather than chemical wet etching \cite{Rub1998} to pre-structure the back-gates needed to control the electron density of the lower 2DEG. The ion implantation technique is also capable of producing many samples simultaneously. It is planar and no wafer space is wasted to accommodate the extremely shallow etched edges of the back-gates. Typical gate voltages are of the order of 1 V which is much smaller than those necessary for gates placed on the back side of the wafer.\cite{Eisenstein1990}

Results on two types of structures are reported. In the first sample, we use a bilayer with a barrier of 6 nm, which results in relatively small tunneling. It is equipped with separate contacts, based on implanted gates and the selective depletion technique. The densities can be varied separately and show an anomalous behavior at equal density because of enhanced  interlayer tunneling.

A second structure contains a barrier of only 3 nm which results in strong tunneling with no contact separation. Due to the planar back-gate design we find none of the problems, which occur frequently with the non-planar wet-etching design. For example the freeze out of the ohmic contacts in higher magnetic fields, which is a result of the poor mobility of the wet-etched samples. This  allows us to map a large range of filling factor combinations and in particular of the transition from the fractional quantum Hall state (FQHS) to the Wigner-crystal phase. Most observed features are a consequence of the single layer properties. Interestingly, we find that the single-layer filling-factor 1/5 is not a FQHS but an insulating phase in our quantum well structures. 

\section{Experimental details}
\begin{figure}[h]
	\includegraphics[width=0.5\textwidth]{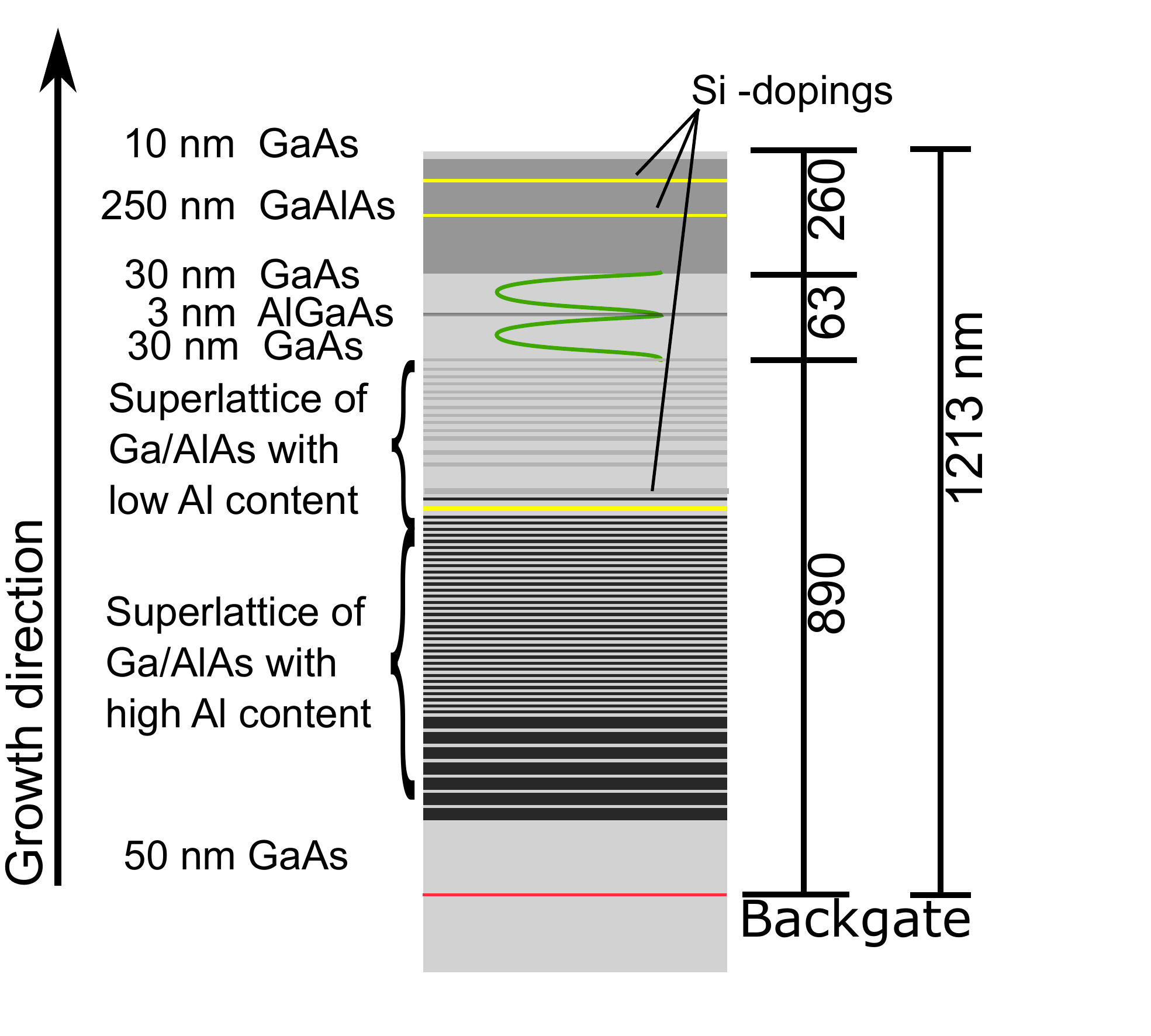}
	\caption{Complete layer stack of the sample, the region of the electron gas is indicated by the two green curves, representing the electron wavefunctions. The yellow lines indicate the doping, the red line the MOCVD grown Si-doped back-gate layer. Super-lattices with high aluminum content are grown below the 2DEG to suppress conduction between back-gate and 2DEG. Close to the quantum wells the aluminum content is drastically reduced to sustain high mobility 2DEGs.}
	\label{fig:layers}
\end{figure}

The samples are produced by overgrowing a processed back-gate GaAs substrate using molecular beam epitaxy (MBE). 
The separate tuning of the electron densities in the upper and lower quantum well requires both top and back-gates. While the preparation and operation of the top gate is rather trivial, structured back-gates are a technological challenge. We use a back-gating design based upon ion implantation prior to MBE growth which was recently introduced.\cite{Berl2016} 
The patterned back-gates are fabricated by oxygen ion implantation\footnote{implanted at IBS, Peynier, France} into a highly silicon doped gallium arsenide layer. This metal organic vapor deposition (MOCVD) grown layer has been masked with photo-resist prior to the implantation. The most important process step is the residueless removal of this photo-resist prior to the MBE growth, using the epi-ready cleaning process described in \cite{Berl2016}. Using molecular beam epitaxy a gallium/aluminum arsenide heterostructure has been grown on top of the wafer. The detailed structure of the strong tunneling sample is shown in Fig. \ref{fig:layers}. 
Having a small overlap between the back-gate and the ohmic contacts is crucial for accumulating the 2DEG because any ungated area can generate measurement artifacts at high magnetic fields. In Fig.~\ref{fig:ImageSC} the back-gate/contact overlap is indicated in red.  
To achieve reliable, leak-proof back-gates it is equally important to have a high aluminum concentration in the buffer structure below the quantum wells. On the other hand the Al concentration is drastically reduced closer to the quantum well to achieve a 2DEG of high mobility. 

A Hall-bar of dimensions 650 $\mu$m $\times$ 200 $\mu$m is prepared using standard photo-lithography and wet etching. An Au/Ge/Ni alloy is deposited and annealed to contact the back-gates and both upper and lower 2DEG in parallel. In the last step the gold top-gates are evaporated onto the Hall-bar. If contact separation is required, additional pinch-off gates are added. These require no additional process steps as the pinch-off gates are prepared together with the global top- and back-gate. A sketch of the structure is shown in Fig.~ \ref{fig:densitydownSC}.  

All measurements are performed in a dilution refrigerator with a base temperature of about 25 mK. The  longitudinal $R_{xx}$ and Hall resistance $R_{xy}$ are measured using standard lock-in techniques.  Individual contact resistances are in the range of a few hundreds Ohms. The current through the Hall-Bar is set between 1 and 10 nA to reduce electron heating at a frequency of 13.8 Hz and is always monitored by another Lock-in amplifier. 

In the following part we will present details and characterization measurements on the weak interlayer tunneling sample with separated contacts for the two 2DEGs, followed by the section on strongly tunneling bilayer systems with parallel contacts.  

\section{Bilayers with weak interlayer tunneling and separate contacts}

\begin{figure}[h]
	\includegraphics[width=0.4\textwidth]{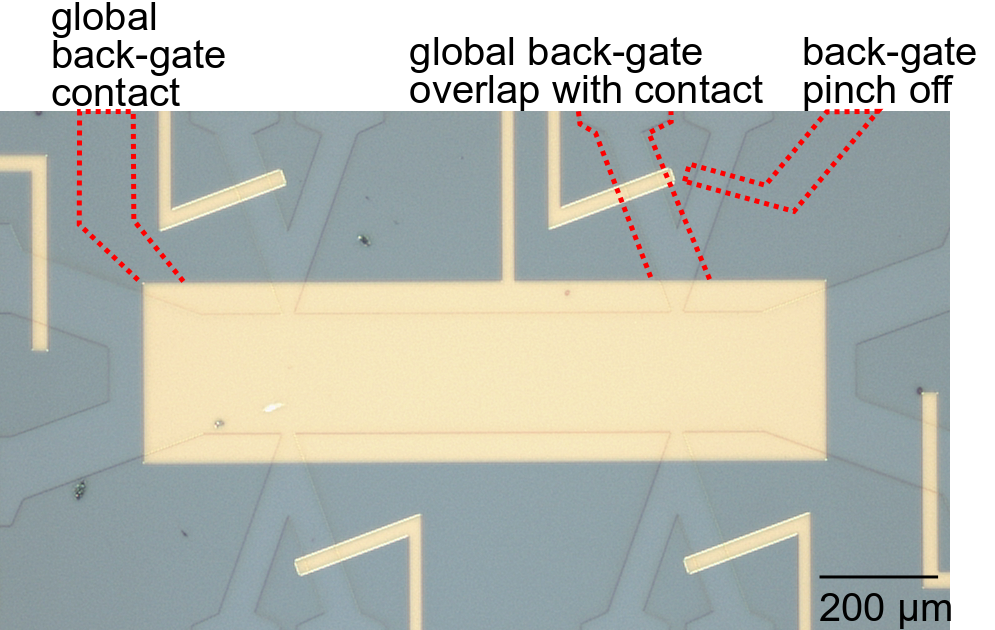}
	\caption{Microscopic image of the Hall-Bar, the thin lines, visible in the structure, marks the etched mesa, in gold the global top- and the pinch-off top-gates are shown. As a guide to the eye, for one contact pair both back-gates are indicated, as well as contact to the global back-gate. All 6 pairs of contacts are equivalent.}
	\label{fig:ImageSC}
\end{figure}

A bilayer with separate contacts requires a sophisticated mask design to produce separate ohmic contacts to each layer, and to control the carrier concentrations individually. In Fig. \ref{fig:ImageSC} we show a microscopic image of a sample with such contact separation. The center part shows the Hall bar with the global top gate made of an Au film. Six V-shaped pairs lead to the Hall bar. One arm of each pair is crossed by an Au film which represents the top pinch-off gate. The back-gates are not visible because the ion implantation does not modify the surface. This type of back-gate does not significantly interfere with the conductivity of the quantum well electrons crossing them, which results in high mobility 2DEGs. For one contact pair the back-gates are outlined in red to clarify the design.

\begin{figure}[h]
	\includegraphics[width=0.5\textwidth]{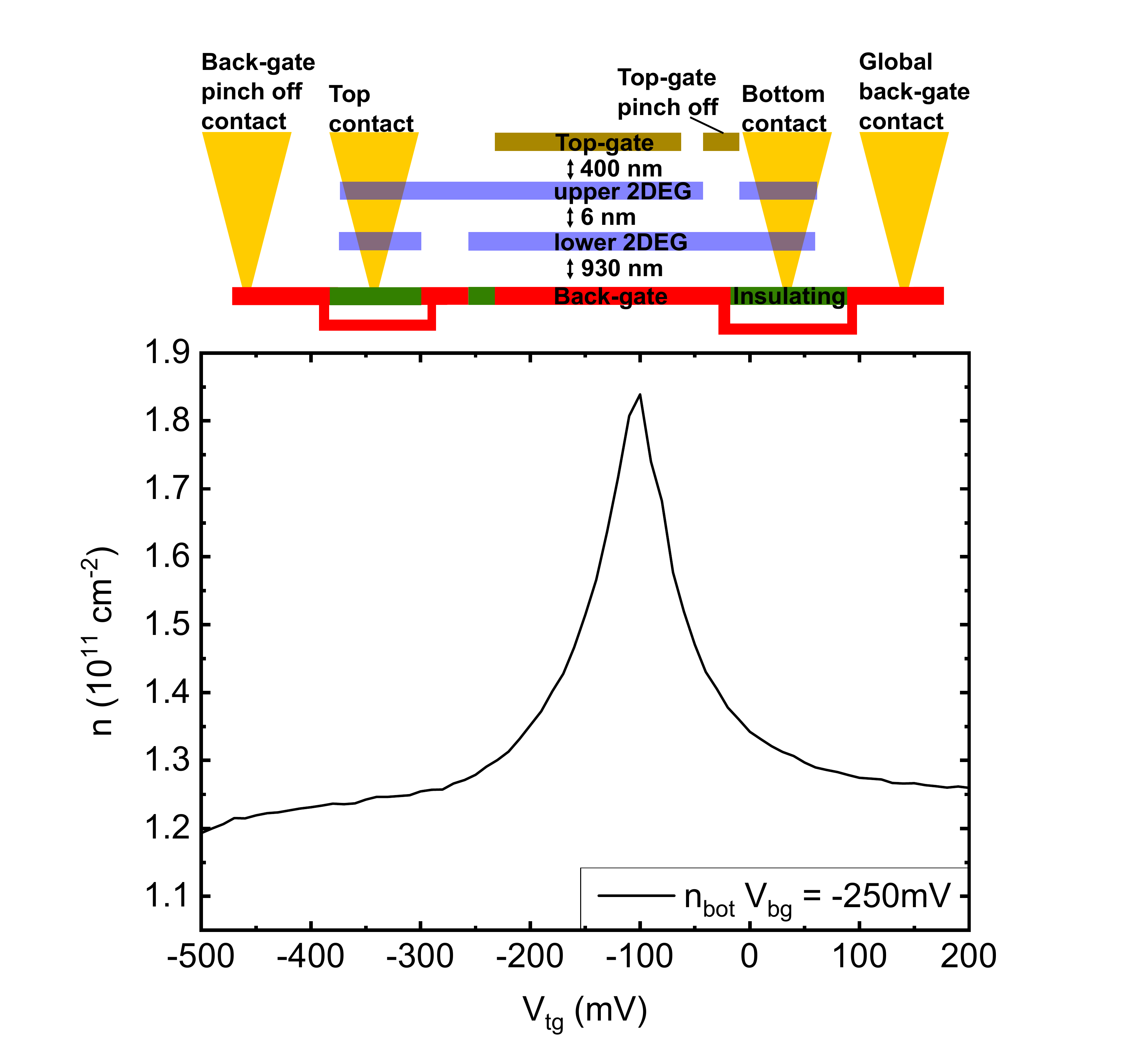}
	\caption{The lower 2DEG density is plotted versus the top-gate voltage. As expected no change in density is observed, however, at equal densities a peak can be observed. The sketch in the upper panel shows the relevant length scales of the sample as well as the operation of the pinch-off gates.}
	\label{fig:densitydownSC}
\end{figure}

The MBE-grown structure is similar to the one shown in Fig. \ref{fig:layers}, except that  the quantum wells are now 18.7 nm wide and separated by a 6 nm Al$_{0.8}$Ga$_{0.2}$As barrier to have reduced tunneling. A schematic of the sample is shown in the sketch in Fig.~\ref{fig:densitydownSC} with the relevant length scales as well as the design of the pinch-off gates. 

Contacts are separated by selectively depleting the 2DEG underneath a top pinch off gate and over a back pinch off gate \cite{Eisenstein1990}\cite{yoon2010}. In a first step the voltage drop between two contacts of the Hall-bar as a function of either pinch-off top- or back-gate voltage is measured. Plateaus in the resistance indicate the depletion of one layer. Depletion of both layers render the system insulating. The voltage values at the plateaus are then applied to all the respective pinch-off gates. For our sample the required pinch-off back-gate voltage is -2 V and for the front gates -0.8 V. 

\begin{figure}[h]
	\includegraphics[width=0.5\textwidth]{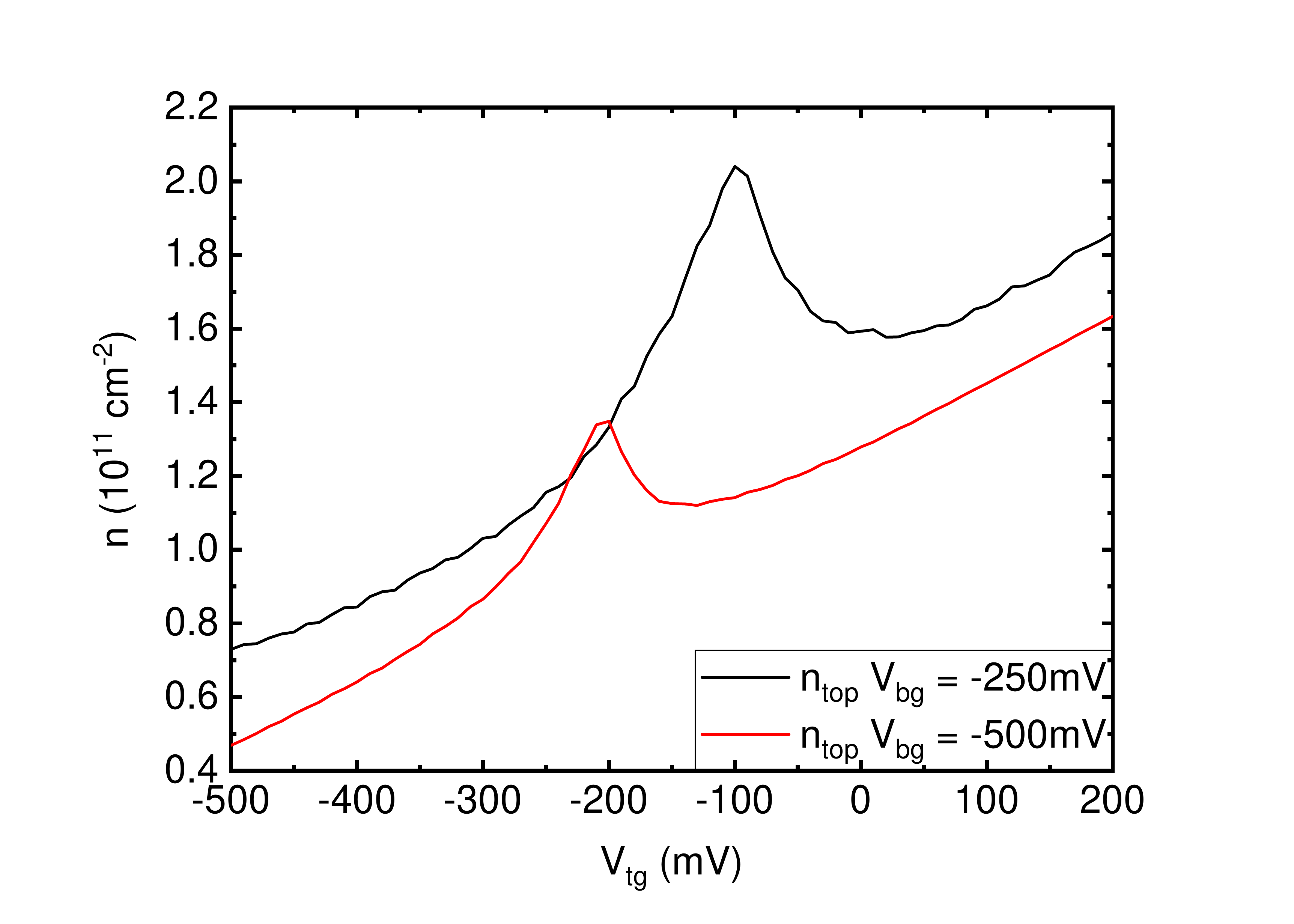}
	\caption{The upper 2DEG density is mapped against the top-gate voltage for two different back-gate voltages. The linear increase in density is interrupted by peaks at equal densities in both layers.}
	\label{fig:densityupSC}
\end{figure}

With the pinch-off gates activated, the two layers can now be characterized separately. As an example, we show the density deduced from the Hall voltage measured at 0.1 Tesla, of the top layer as function of the global top gate voltage for two different back-gate voltages in Fig.~\ref{fig:densityupSC}.  A general increase of the density with voltage is observed as expected. However, there are pronounced maxima at -200 mV and -100 mV top gate voltage, respectively. These maxima seem to signal that charges of the other layer contribute to the density measured via the Hall effect. Indeed, if we measure the density in the lower layer again as a function of the top-gate voltage we find a similar peak at the same voltage of -100 mV which is on top of a rather constant lower layer density, see Fig.~\ref{fig:densitydownSC}.

We interpret these peaks as a consequence of a resonant tunneling phenomenon which occurs at equal densities in the two layers. Under this condition the Fermi wave vectors in the two layers are equal and the in-plane wave vectors of the tunneling electrons are conserved,  consequently the tunneling is enhanced. Due to the strong wave function overlap, the bilayer system can be viewed as a single (wide) quantum well with almost doubled density. This resonance phenomena has the same origin as the one  observed earlier in the tunnel characteristics of bilayer systems.\cite{Eisenstein1991} 


The strong tunneling phenomena encountered here demonstrate that bilayers cannot be operated separately using even thinner barriers. Consequently, the 3 nm barrier sample which we describe in the next section is produced without the separate contacting features. With that sample we investigate the influence on tunneling on the magneto transport of bilayers in the regime of the fractional quantum Hall effect and the metal-insulator transition.

\section{Bilayers with strong interlayer tunneling without separate contacts}

\begin{figure}[h]
	\includegraphics[width=0.5\textwidth]{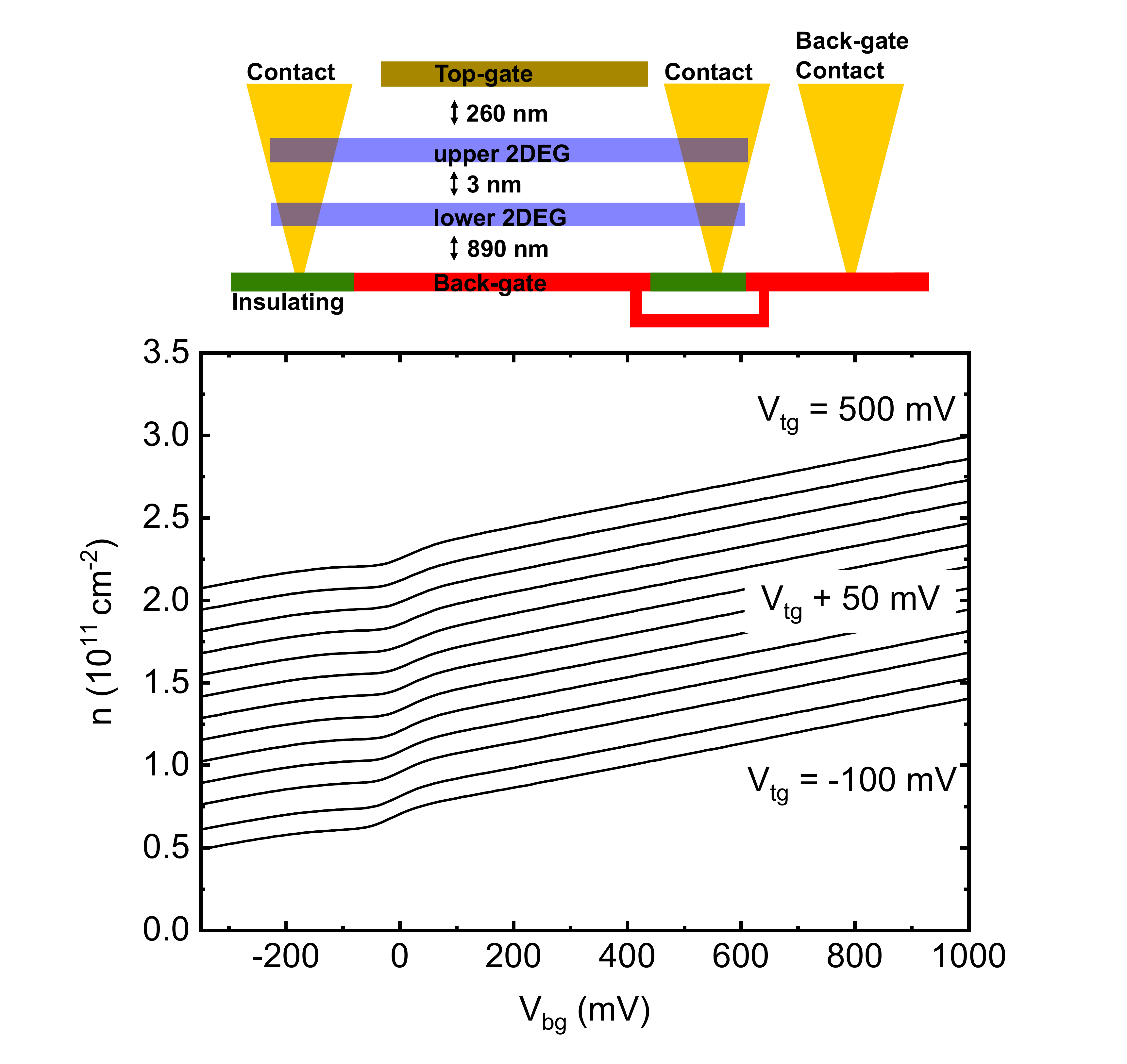}
	\caption{Density as function of the back-gate voltage for different top gate voltages. The kink in the density at around V$_{bg}$ = -100 mV  is due to the transition from a single to a bilayer system. The upper panel illustrates the spatial separation between the gates and the 2DEGs.}
	\label{fig:density}
\end{figure}

The layer structure is similar to the one of the previous section. Both  GaAs quantum wells are again 30 nm wide, but the width of the Al$_{0.27}$Ga$_{0.73}$As barrier is only 3 nm. The contact pinch-off gates are not necessary and are omitted, see in the sketch of Fig.~ \ref{fig:density}. 

\begin{figure}[h]
	\includegraphics[width=0.5\textwidth]{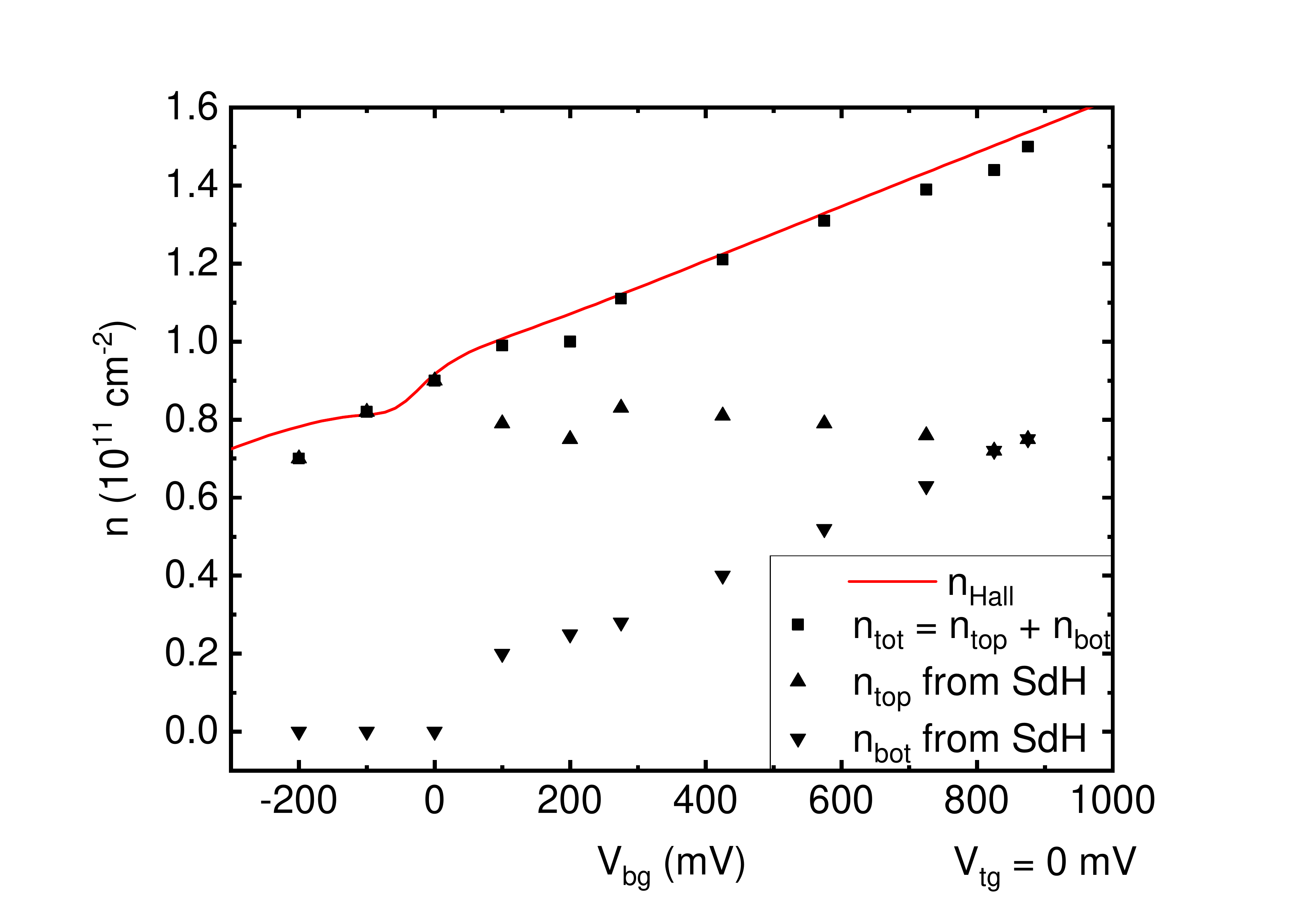}
	\caption{Comparing the density from the Hall measurement with densities derived from Shubnikov de Haas oscillations at larger magnetic fields for V$_{tg}$ = 0 mV.  
	}
	\label{fig:separate_density}
\end{figure}

Figure~\ref{fig:density} shows the nearly linear increase of the total density as a function of back-gate voltage for a series of top-gate voltages. The kink at around V$_{bg}$ = -0.1V is due to the formation of the second (bottom) 2DEG. At higher density, the original density versus back-gate behavior is recovered. In this bilayer regime the slope is slightly larger due to the difference in the back-gate to quantum well separation. Although there are no separate contacts, one can deduce the individual densities in the two layers from the magnetotransport experiments which will be shown later. For V$_{tg}$ = 0 V, Fig. \ref{fig:separate_density} compares the carrier densities deduced from the Hall voltage with densities deduced from the magnetic field position of integer and fractional quantum Hall states. The cumulated density matches those measured by the Hall voltage which demonstrates the reliability of the gating.      

\begin{figure}[h]
	\includegraphics[width=0.45\textwidth]{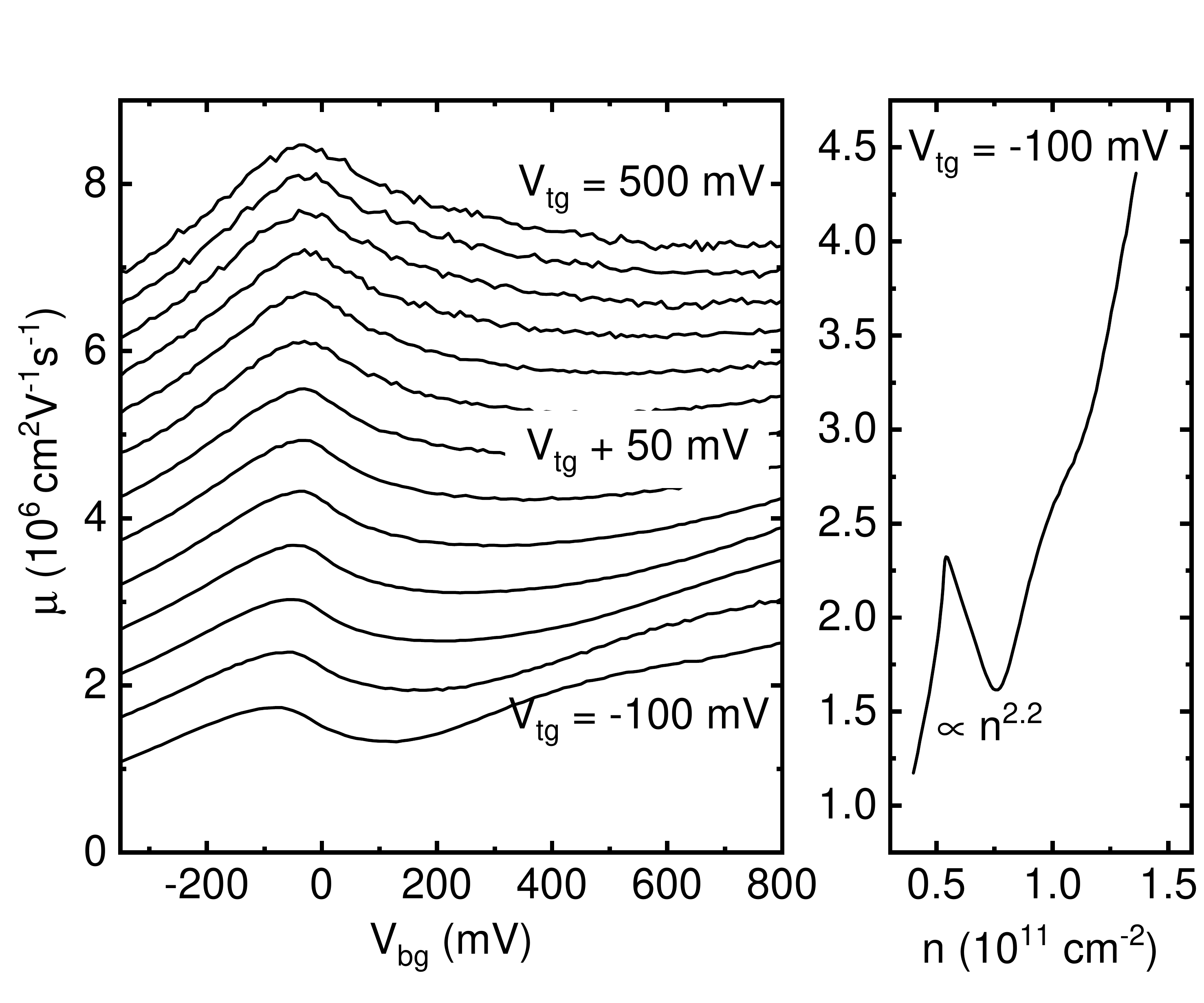}
	\caption{Left: Mobility as a function of the back-gate voltage for different top gate voltages. From V$_{bg}$ = -100 mV upwards it is the combined mobility for the two layers, below it is the single layer mobility, corresponding to figure \ref{fig:density}. Right: The mobility is plotted as a function of the density for a V$_{tg}$ = -100 mV. The slope for the single layer regime is indicated. Up to a density n =  0.55 $\times$ 10$^{11}$ cm$^{-2}$, only the upper layer is populated with electrons.}
	\label{fig:mobility}\label{fig:alpha}
\end{figure}

The mobility measured in the bilayer is shown in Fig.~\ref{fig:mobility}. The maximum electron mobility reaches 8 $\times 10^{6}$  cm$^2$V$^{-1}$s$^{-1}$ at an upper layer density of about 2.3 ${\times} 10^{11}$  cm$^{-2} $.
Most experiments were done with V$_{tg}$= 0 V where the mobility is about 3 $\times 10^{6}$  cm$^2$V$^{-1}$s$^{-1} $, it drops slightly as the second layer forms but remains above  2 $\times 10^{6}$  cm$^2$V$^{-1}$s$^{-1} $ for all  V$_{bg}$ values. The mobility of the bottom layer is comparable to that of the top layer. By populating only a single layer, we are able to deduce the individual mobilities.

Interestingly, the mobility of the first layer increases much more rapidly with density than expected. An example is shown in Fig.\ref{fig:alpha} where the mobility $\mu$ is plotted vs. the combined bilayer density \textit{n} for V$_{tg}$= -100 mV. In the initial single layer regime the mobility increases with \textit{n}$^{2.2}$, the exponent is much larger than the usual 0.7 to 1.0 values expected for single layers. Similar features have been previously discussed by Jiang et al.\cite{jiang1988}, where they assume that for very low  \textit{n}, density fluctuations explain the larger exponent. Also striking is the rapid decrease of the mobility with increasing total charge after the formation of the second layer. 


\section{High magnetic field measurements}
The extremely high quality of these bilayer samples should permit the emergence of various exotic quantum states at high magnetic fields. By independently varying the carrier densities in both layers, we have studied a large parameter space and different regimes of interlayer interactions.


The top-gate voltage V$_{tg}$ sets the starting density in the upper layer. With this voltage kept fixed the longitudinal \textit{R}$_{xx}$ and the Hall resistance  \textit{R}$_{xy}$ of our structure is measured as a function of back-gate voltage and magnetic field. 
The resistivity  $\rho_{xx} $ is obtained by scaling  \textit{R}$_{xx}$ with the length 650 $\mu$m to width 200 $\mu$m ratio and is represented as a two-dimensional color rendering  as a function of  \textit{V}$_{bg}$ and magnetic field  \textit{B}.

\begin{figure}[h]
	\includegraphics[width=0.5\textwidth]{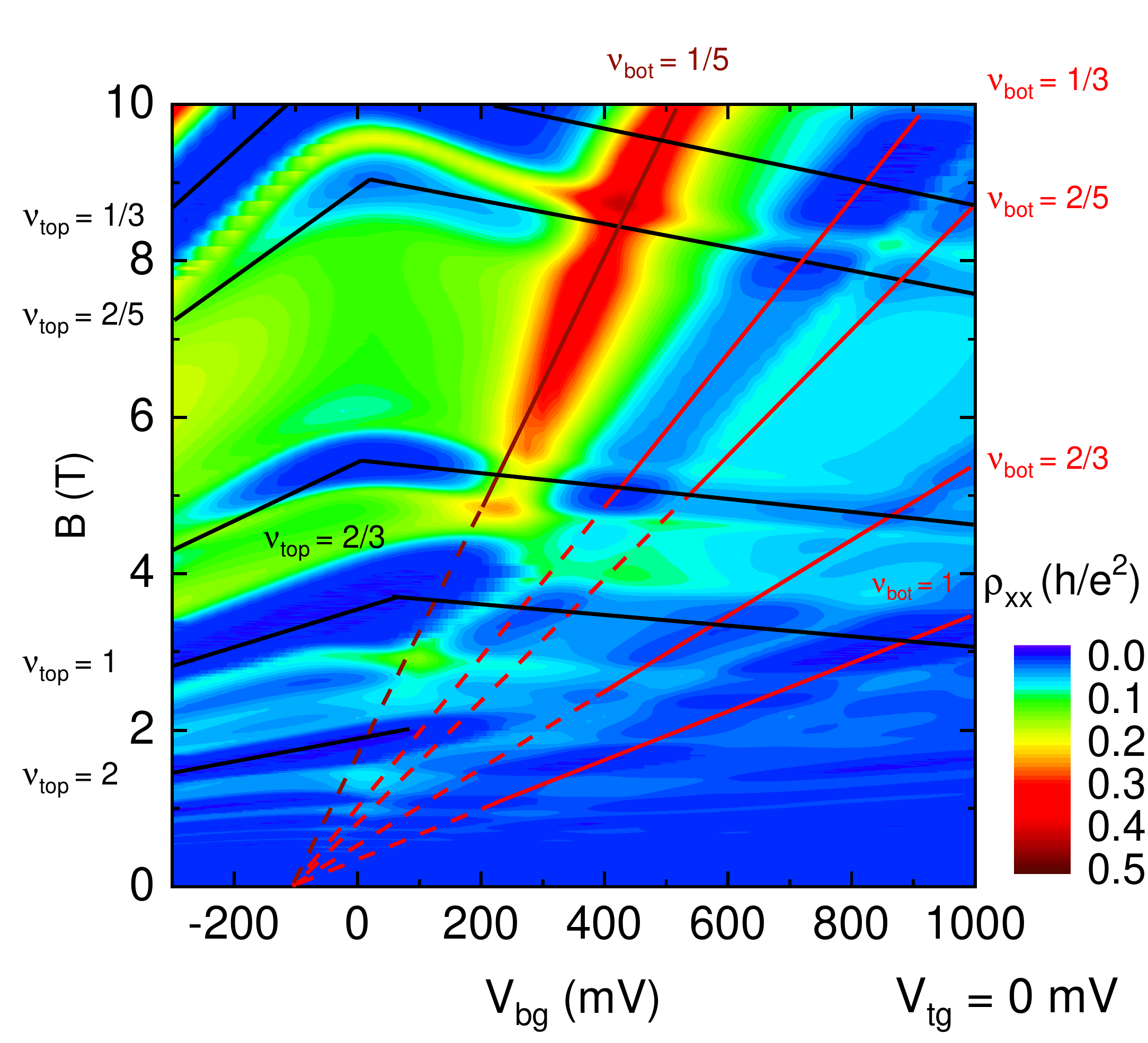}
	\caption{Color plot of the longitudinal resistivity at 0 mV top-gate voltage. Individual filling factors are indicated for each layer, black for the upper , red for the lower 2DEG. Additionally, an unexpected region of high resistivity is marked which extends approximately along the bottom filling factor 1/5.}
	\label{fig:phase1}
\end{figure}

For  \textit{V}$_{bg}$ less than -100 mV the lower quantum well is empty (cf. Fig.~\ref{fig:density}) and does not screen the electric field of the back-gate. Thus, the density of the upper quantum well increases with increasing  \textit{V}$_{bg}$.  This leads to a Landau-Fan typical for a single 2DEG in the left part of the color rendering of Fig.~\ref{fig:phase1}. The blue areas in the plot denote $\rho_{xx}$ values near zero Ohm. Colors from green to red mark larger resistances. Particularly significant in the Landau fan are the $\nu$=1/3  fractional and the $\nu$=1 integer QHE states. These and the other significant fractions are labeled with black lines in the Figure. The Hall resistance measured simultaneously facilitates the identification of the filling factors. The area of filling factor 2/3 is interrupted at smaller fields, most likely due to the transition of the spin-unpolarized to a polarized ground state.\cite{Chakra1990}

As soon as the back-gate voltage exceeds -100 mV, electrons start to accumulate in the lower layer. The position in magnetic field of the QHE states in the top layer are not completely independent from the back-gate as one would have expected. Instead one observes that the top layer QHE states shift to lower magnetic fields indicating a decrease of density in the top layer. This is a consequence of negative compressibility of the lower layer at small densities.\cite{Eisenstein_1992,Ying1995,Zheng1997,Zhang_2013} This behavior is consistent with the density dependence shown in Fig. \ref{fig:separate_density}. 


With further increasing \textit{V}$_{bg}$ a second Landau fan develops with a similar series of QHE states. The corresponding filling factors from the lower layer are labeled with red lines in the Figure. 
Resistivities of two parallel layers in high magnetic fields can be estimated by inverting the sum of the respective conductivity tensors.\cite{Grayson2006} This leads to a zero longitudinal resistivity only if both layers are in a QHE state. A resistivity minimum is found if one of the two layers is in a QHE state. The Hall resistance of two parallel QHE states is also quantized and corresponds to the sum of the two respective filling factors. For example at the crossing of the 2/3 (top) and 1/3 (bottom) states we measure a Hall plateau corresponding to filling factor 1 while at the crossing of 1/3 (top) and 2/5 (bottom) the plateau value signals a 11/15 filling which appears unusual but is a rather trivial consequence of the bilayer system. Inspecting Fig.~\ref{fig:phase1} shows that the combined resistivity of a bilayer follows this rule at least qualitatively for most combinations of QHE states with a non-QHE state.

\begin{figure}[h]
	\includegraphics[width=0.5\textwidth]{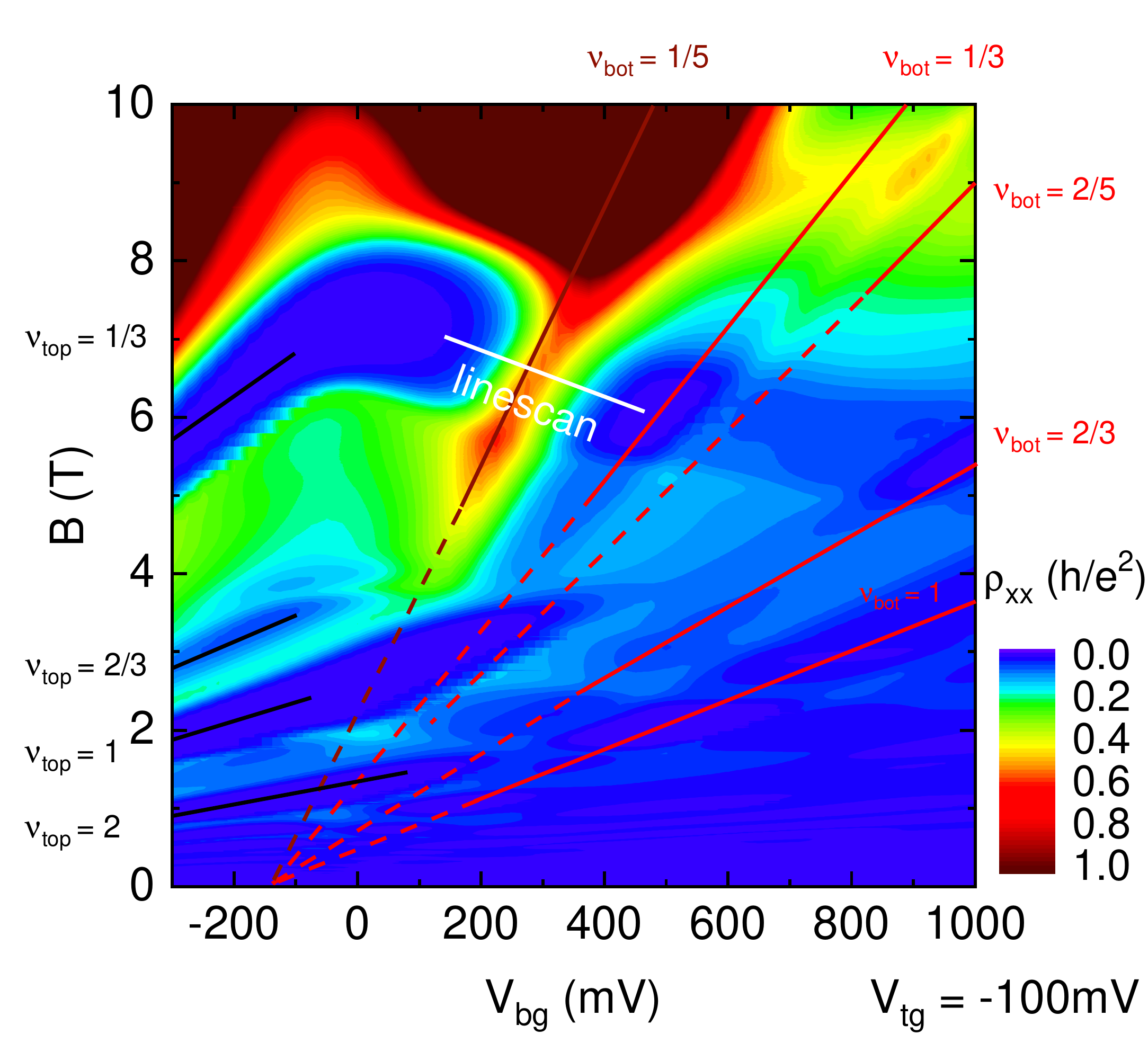}
	\caption{Color plot with -100 mV applied to the top gate. The extracted line scan, shown in figure \ref{fig:Linescan} is indicated. }
	\label{fig:phase2}
\end{figure}

\begin{figure}[h]
	\includegraphics[width=0.5\textwidth]{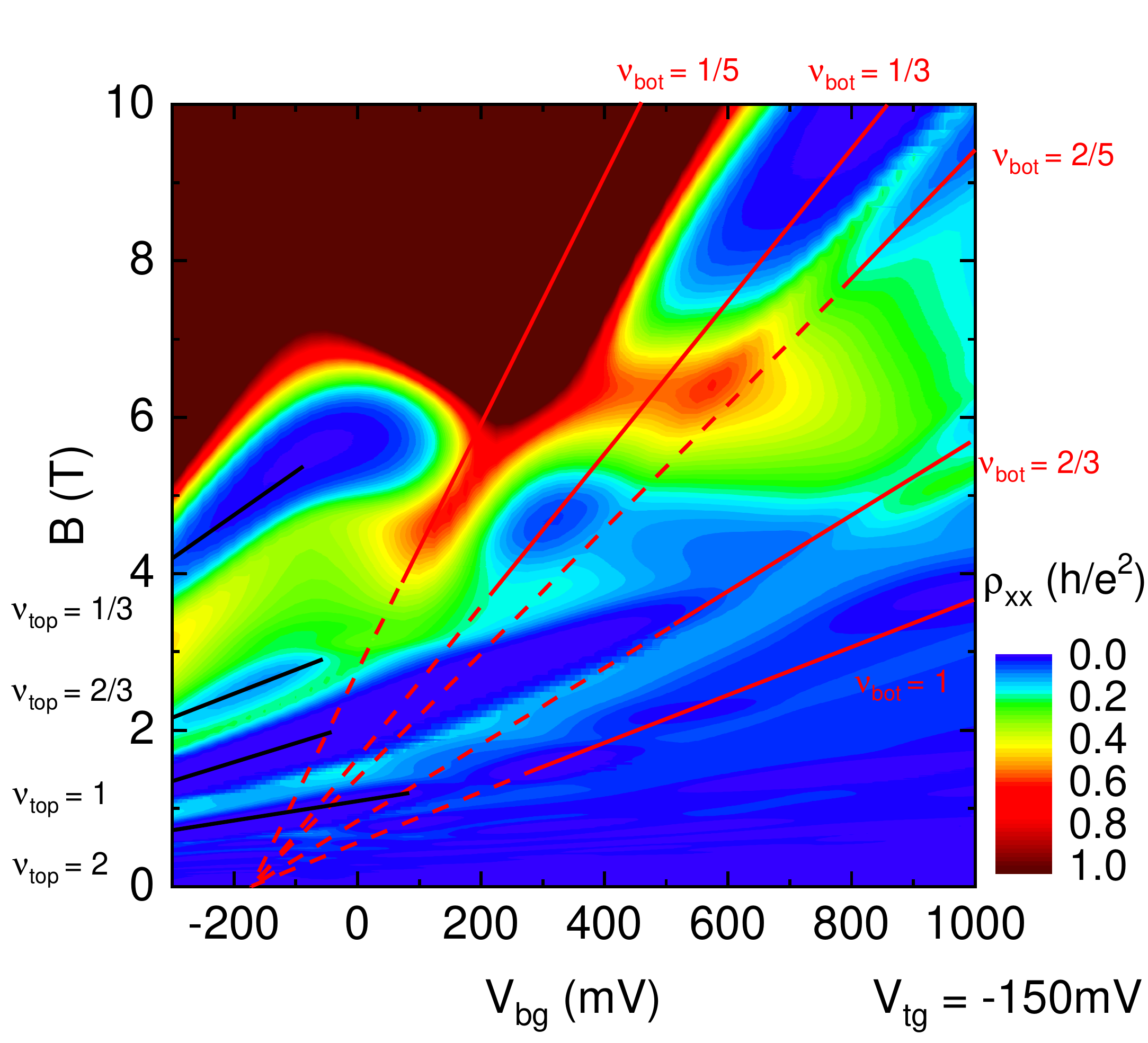}
	\caption{Color plot with -150 mV applied to the top gate.}
	\label{fig:phase3}
\end{figure}

In Figures~\ref{fig:phase2} and \ref{fig:phase3} we present results for top gate voltages of -100 mV and -150 mV, respectively, which reduce the density of the top layer. The overall reduction of the carrier concentration leads to insulating behavior at large magnetic fields and low densities (small back-gate voltages) in the lower layer.

Most features in Figures~\ref{fig:phase1},\ref{fig:phase2} and \ref{fig:phase3} can be accounted for by known QHE states of single layers. One exception is the excitonic BCS-like QHE state at total filling one. It starts at filling one in the single top layer, and extends for decreasing top-gate voltages. For \textit{V}$_{tg}$ = 0 V, it only exists until the lower layer becomes populated. For decreasing densities in the top-layer, it stretches until around \textit{V}$_{bg}$ = 400 mV in Fig.~\ref{fig:phase2} and even further to \textit{V}$_{bg}$ = 600 mV in Fig.~\ref{fig:phase3}. It can also be identified by the slightly different slope compared to the single layer $\nu_{top}$ = 1.

\begin{figure}[h]
	
	\includegraphics[width=0.5\textwidth]{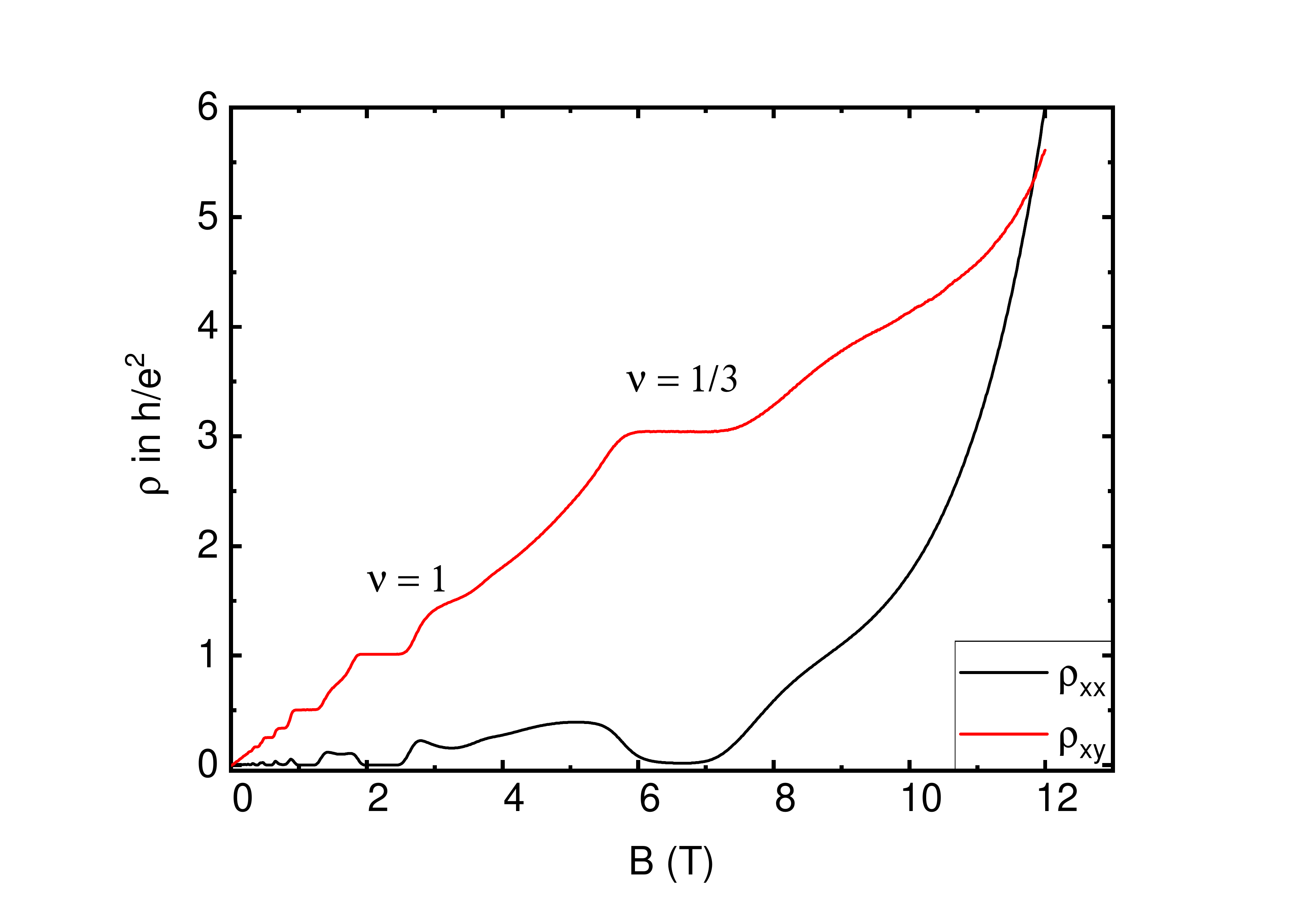}
	\caption{Magnetotransport as a function of magnetic field for V$_{tg}$ = -100 mV and V$_{bg}$ = -310 mV. No FQHS corresponding to filling factors smaller than 1/3 can be observed.}
	\label{fig:Bscan}
\end{figure}

The other unexpected feature is the maximal resistivities around filling factor $\nu_{bot}$ = 1/5 seen in all three color plots. This indicates that the filling factor 1/5 is not an FQHE state in our system contrary to expectation. This is, however, not a bilayer phenomena as the inspection of the Landau fans in the left parts of Figures~\ref{fig:phase2} and \ref{fig:phase3} reveals. It shows that the 1/5 FQHE state does not appear in the single top layer either. For V$_{bg}$ $<$ -150 mV only the top layer is populated and the system should behave as a single layer. The absence of the 1/5 FQHE state is more clearly seen in a sweep of the magnetic field at V$_{tg}$= -100 mV and  V$_{bg}$ = -310 mV which is shown in Fig.~\ref{fig:Bscan}. At fields beyond the filling factor 1/3 the longitudinal resistivity $  \rho_{xx}$  increases rapidly. The increase appears like the metal-insulator transition which is usually only seen beyond 1/5 filling.\cite{Mendez1983,Jiang_1991}

\begin{figure}[h]
	\includegraphics[width=0.5\textwidth]{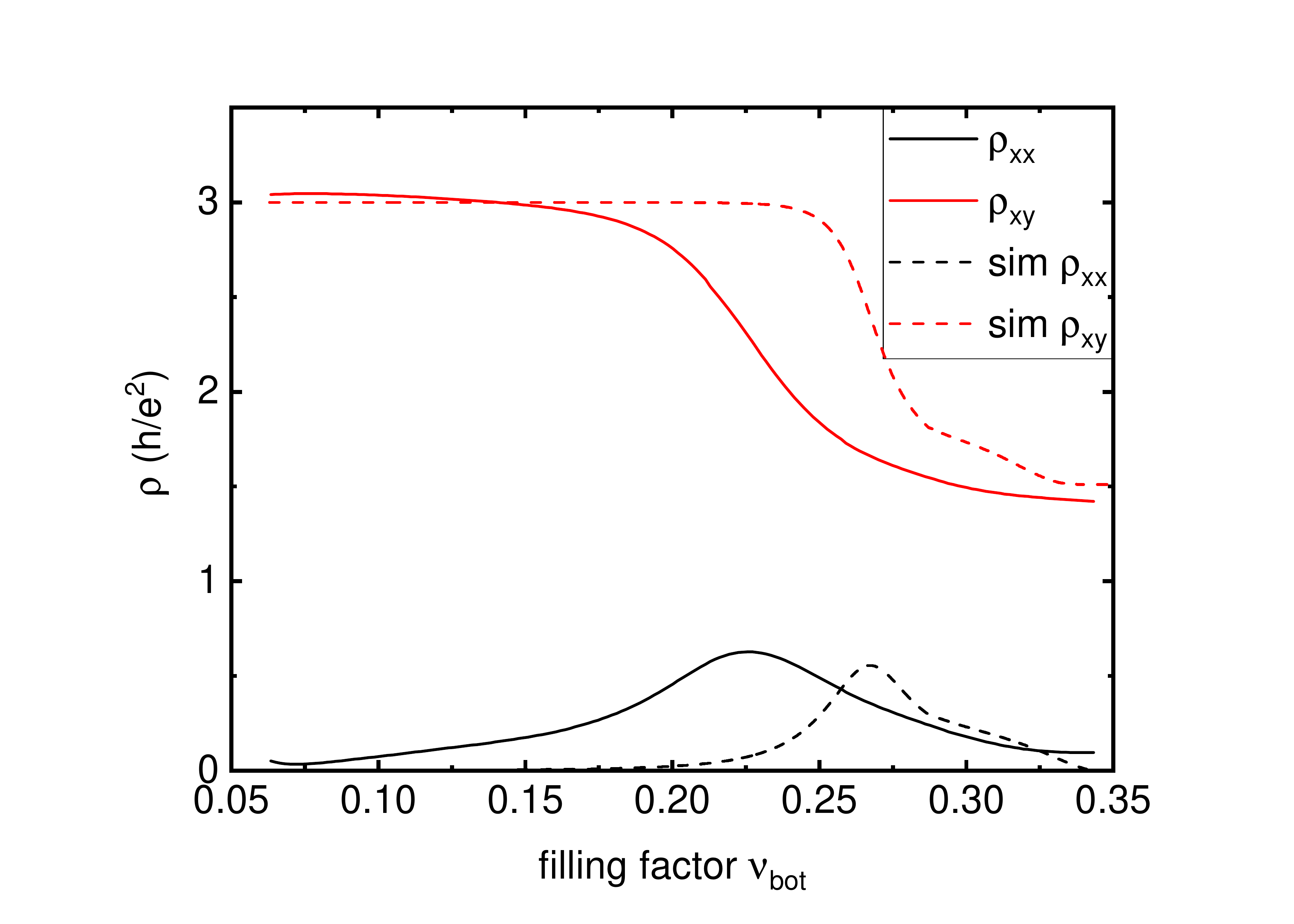}
	\caption{Transport measurement along the line indicated in Fig.~\ref{fig:phase2}, in black and the parallel conductance simulation for one layer fixed at filling factor 1/3 and the high field data, above 6.5 T, from figure \ref{fig:Bscan} in red.}
	\label{fig:Linescan}
\end{figure}

The observation that a single 2DEG layer turns into an insulating state just below 1/3 filling explains the high resistive states along the 1/5 filling factor in the Figs.~\ref{fig:phase1},\ref{fig:phase2} and \ref{fig:phase3}. We now focus in Fig.~\ref{fig:phase2} on the white line, which represents a constant filling factor of $\nu$ = 1/3 in the upper layer but a varying filling factor between 1/15 and 1/3 in the bottom layer. Figure \ref{fig:Linescan} plots the resulting $\rho_{xx} $ and $\rho_{xy}$ along this line. We find that $ \rho_{xx} $ shows a maximum at the center of the transition while the $ \rho_{xy}$ curve shows a smooth transition from the 1/3 quantized value to 2/3 filling. 

The behavior of  $ \rho_{xx} $ and $ \rho_{xy}$ can be numerically simulated by assuming that the top layer remains at 1/3 filling with $ \rho_{xx} $ and $ \rho_{xy} $ set to zero and 77.4 $ k\Omega $, respectively. For the bottom layer we use the  $ \rho_{xx} $ and $ \rho_{xy} $ data of Fig.~\ref{fig:Bscan} for the range of filling factor 1/3 and below, plotted as a function of filling factor rather than magnetic field. By extrapolation of $ \rho_{xx} $ and $ \rho_{xy} $, we expand the measurement data down to a density corresponding to filling factor 1/15 in the bottom layer. Again using the procedure of summing the conductivity tensors we simulated $ \rho_{xx} $ and $ \rho_{xy}$ values of the bilayer and plot them also in Fig.~\ref{fig:Linescan}. Both simulated resistivity components show qualitatively a very similar behavior as the measured ones. A peak in  $ \rho_{xx} $ occurs for both experiment and simulation near filling factor 1/5 with approximately the same height while $ \rho_{xy}$ shows a smooth transition from the single layer 1/3 to the bilayer 2/3 state. The main difference is that the measured increase in the longitudinal resistance seems to be smaller than for the expanded data used in the simulation, as the simulated peak is shifted to larger filling factors. 

\section{Conclusions}

We successfully demonstrated that fully controllable bilayer systems can be prepared by using ion implantation for pre-structuring the back-gates. The first type of sample presented here contains pinch-off gates in addition to the global front and back-gates allowing the separate contacting and characterizing of the two layers. It turns out that the layer separation breaks down at equal densities because of the enhanced tunneling which occurs at coinciding Fermi wave-vectors. 


Magneto transport measurements of a second bilayer sample without layer separation at high
magnetic field allowed not only to investigate the metal-insulator transition from a single to a bilayer
system but also to identify novel quantum states. We found that both the metal insulator transitions and the quantum Hall states are dominated by the single layer behavior. Thus, the strongly enhanced tunneling of the 3 nm thick barrier does not lead to novel quantum phenomena in the parameter range covered here.

The excitonic BCS state at the total filling factor one has been reproduced. This state originates from the interlayer Coulomb interaction and has also been observed at much larger barriers.\cite{Wiersma_2006}  

Noteworthy is the absence of a FQH state at filling factor 1/5 in our samples. Most earlier experiments which observed the filling factor 1/5 FQH state on electrons have been performed on single heterointerface structures, \cite{Jiang_1991,Buhmann1991,Willett1988} while later studies have been on much wider quantum wells.\cite{drichko2015,Pan2002} 
Thus, it appears as if the confinement to a 30 nm wide quantum well has an effect on the metal-insulator transition below 1/3 filling. This deserves further investigation. It seems that the 1/5 FQHE state is very fragile as is also demonstrated by its sensitivity to a screening gate\cite{Deng_2018} and the complicated behavior observed with 2D hole gases.\cite{Zhang2015}

\ack{We acknowledge the support through the Swiss National Foundation (SNF) and the NCCR QSIT (National Center of Competence in Research - Quantum Science and Technology). We are thankful to E. Gini for MOCVD growth in preparation of the back-gates performed at FIRST - Center for Micro- and Nanoscience, ETH Zurich. We thank H. Karl for his insights on the ion implantation technique.}
\newpage
\bibliographystyle{unsrt}
\bibliography{Bilayer_bib}

\end{document}